\documentclass[12pt]{article}
\pdfoutput=1 
\usepackage{amsmath}
\usepackage{amsfonts}
\usepackage{algorithm}
\usepackage{algorithmic}
\usepackage{natbib}
\usepackage{color}
\usepackage{caption}
\usepackage{subcaption}
\usepackage{graphicx}
\usepackage{url}
\usepackage{hyperref}
\usepackage{etoolbox}
\usepackage{authblk}

\makeatletter

\patchcmd{\NAT@citex}
{\@citea\NAT@hyper@{%
		\NAT@nmfmt{\NAT@nm}%
		\hyper@natlinkbreak{\NAT@aysep\NAT@spacechar}{\@citeb\@extra@b@citeb}%
		\NAT@date}}
{\@citea\NAT@nmfmt{\NAT@nm}%
	\NAT@aysep\NAT@spacechar\NAT@hyper@{\NAT@date}}{}{}

\patchcmd{\NAT@citex}
{\@citea\NAT@hyper@{%
		\NAT@nmfmt{\NAT@nm}%
		\hyper@natlinkbreak{\NAT@spacechar\NAT@@open\if*#1*\else#1\NAT@spacechar\fi}%
		{\@citeb\@extra@b@citeb}%
		\NAT@date}}
{\@citea\NAT@nmfmt{\NAT@nm}%
	\NAT@spacechar\NAT@@open\if*#1*\else#1\NAT@spacechar\fi\NAT@hyper@{\NAT@date}}
{}{}

\newcommand{\blind}{0}
\newcommand*\samethanks[1][\value{footnote}]{\footnotemark[#1]}

\begin{document}
	
	\def\spacingset#1{\renewcommand{\baselinestretch}%
		{#1}\small\normalsize} \spacingset{1}
	
	
	\if0\blind
	{
		\title{\bf Persistence Terrace for Topological Inference of Point Cloud Data}
		\author[1]{Chul Moon\thanks{Supported in part by NSF IIS-1607919.}}
		\author[2]{Noah Giansiracusa\thanks{Supported in part by NSA grant H98230-16-1-0015.}}
		\author[1]{Nicole A. Lazar\samethanks[1]}
		\affil[1]{University of Georgia}
		\affil[2]{Swarthmore College}
		\date{}
		\maketitle
	} \fi
	
	\if1\blind
	{
		\bigskip
		\bigskip
		\bigskip
		\begin{center}
			{\LARGE\bf Persistence Terrace for Topological Inference of Point Cloud Data}
		\end{center}
		\medskip
	} \fi
	
	\bigskip
	\begin{abstract}
Topological data analysis (TDA) is a rapidly developing collection of  methods for studying the shape of point cloud and other data types.  One popular approach, designed to be robust to noise and outliers, is to first use a smoothing function to convert the point cloud into a manifold and then apply persistent homology to a Morse filtration.  A significant challenge is that this smoothing process involves the choice of a parameter and persistent homology is highly sensitive to that choice; moreover, important scale information is lost.  We propose a novel topological summary plot, called a persistence terrace, that incorporates a wide range of smoothing parameters and is robust, multi-scale, and parameter-free.  This plot allows one to isolate distinct topological signals that may have merged for any fixed value of the smoothing parameter, and it also allows one to infer the size and point density of the topological features. We illustrate our method in some simple settings where noise is a serious issue for existing frameworks and then we apply it to a real data set by counting muscle fibers in a cross-sectional image.
	\end{abstract}
	
	\noindent%
	\vfill
	
	\newpage
	\spacingset{1.45} 
	\section{Introduction}
	\label{sec:int}

A defining characteristic of many modern data applications -- whether they fall
under the heading of ``Big Data'' or not -- is their unstructured nature.  It can no longer be assumed that data will come to us for analysis in regular
arrays with fixed numbers of rows and columns and a single observation in each cell, usually representing the measured value of a particular variable for a particular unit.  Similarly, questions of scientific interest have been shifting in recent years.  In settings such as neuroimaging and genetics, to
give two prominent examples, researchers are focusing on questions about network
structure, interactions between brain regions or regions on the genome, and the
like.  Such questions are not amenable to traditional statistical procedures based on simple array-structured data.  Accordingly, recent years have seen the development of methods for functional \citep{Ramsey2005}, object-oriented \citep{Marron2014}, and symbolic \citep{Billard2007} data. All of these aim to tackle situations where the basic unit of analysis is something other than a traditional observation; rather, it can now be an entire image, or a histogram, or a function.  

A relatively new, emergent approach is topological data analysis (TDA), which focuses on the ``shape'' of a data set  \citep{Carlsson2009, Carlsson-nature}.  A popular source of data for successful applications of TDA has been biology; examples include the study of brain artery structure \citep{Bendich2016}; brain networks \citep{Lee2012, Petri2014}; protein structure \citep{Gameiro2015}; viral genomics \citep{viral-topology}; bone structures \citep{Heo2012, Turner2014}; and breast cancer \citep{Cancer}.

A prominent branch of TDA is persistent homology, which analyzes the dynamics of the topological features of a data set.  Features are defined through a filtration; as the filtration value changes, topological features are ``born'' (appear) and ``die'' (disappear).  A major challenge in the TDA approach is to decide which features are real, in the sense of representing meaningful structure in the data, and which are artifacts of the noise inherent in all measured data.  This is a question of statistical inference, and as such, recent years have seen the development of statistical methods to blend with the topological concepts underlying persistent homology.  \cite{Fasy2014} determine significant features using a confidence set for the persistence diagram, which summarizes the persistent homology of a data set.  Another confidence set approach \citep{Chazal2015} is based on the persistence landscape, an alternative summary of persistent homology information \citep{Bubenik2015}.  

Persistent homology results are sensitive to noise and outliers in the data, hence robust smoothing-based approaches have also been proposed.  Some works in this direction include \cite{Bobrowski2017}, \cite{Bubenik2015}, and \cite{Phillips2013}.  However, smoothing to make a more robust procedure is not without drawbacks.  First, there might not be a single optimal value of the smoothing parameter, since persistent homology simultaneously tracks topological features appearing in the data at different scales. Second, the smoothing operation emphasizes point density and thereby loses important information about the size of topological features in the data. 

In this paper we suggest a new summary plot that we call the ``persistence
terrace'', due to its three-dimensional staircase structure, that aims to rectify the two disadvantages of smoothing in this context.  The persistence terrace incorporates multiple filtration values (which allows one
to keep track of the births and deaths of topological features) and multiple smoothing parameter values (which removes the need to pick a single smoothing parameter to represent all features), thereby revealing the topological interaction between these two parameters.  In doing so, the persistence terrace helps to distinguish statistical signals that may be overlooked by currently existing methods; moreover, it simultaneously reflects both the size and density of topological features in the data.  We also provide the terrace area plot which displays the areas of every terrace height to help interpret the persistence terrace.  We illustrate our methods on both simulated and real data and offer insight into the meaning of different terrace features, shapes, and sizes.

The rest of paper is organized as follows. In Section 2 we provide mathematical background, then review the standard persistent homology framework; we also discuss limitations when applying it to data analysis.  Section 3 introduces the persistence terrace and demonstrates its important characteristics with
simple examples.  In Section 4 we explore the persistence terrace on more complex simulated and real data.  Finally, in Section 5 we discuss our main findings and give directions for future work.

	\section{Persistent Homology in TDA}
	This section summarizes the theoretical background of (persistent) homology \citep{Hatcher2002,Edelsbrunner2008,Edelsbrunner2010,Ghrist2008,Zomorodian2012} and discusses some statistical shortcomings when applied to data analysis.
		
	\subsection{Homology and Simplicial Complexes}
	A topological space is a mathematical object abstracting the intuitive notions of nearness, connectivity, and continuity. Any metric space (such as $\mathbb{R}^n$), or subset thereof, can be viewed as a topological space.  Homology is used to quantify the topological characteristics of a topological space.  The dimension of the $k$th homology vector space is called the $k$th Betti number, and it counts the number of $k$-dimensional holes in the space.  Table \ref{table:betti} gives the geometric interpretations of the $k=0,1,2$ Betti numbers.
	\begin{table}[!ht]
		\centering    
		\begin{tabular}{|c|c|c|}
			\hline
			\rule{0pt}{2.5ex}
			Symbol & Dimension & Counts \\
			\hline
			\rule{0pt}{2.5ex}
			$\beta_0$ & 0 & Number of connected components \\ \rule{0pt}{2.5ex}
			$\beta_1$ &1 & Number of loops \\ \rule{0pt}{2.5ex}
			$\beta_2$ &2 & Number of enclosed solid voids \\ 
			\hline
		\end{tabular}
		\caption{Interpretation of Betti numbers of dimension zero, one and two ($\beta_0$, $\beta_1$ and $\beta_2$).}
		\label{table:betti}
	\end{table}
	
\begin{figure}[!ht]
		\centering
		\includegraphics[width=15cm]{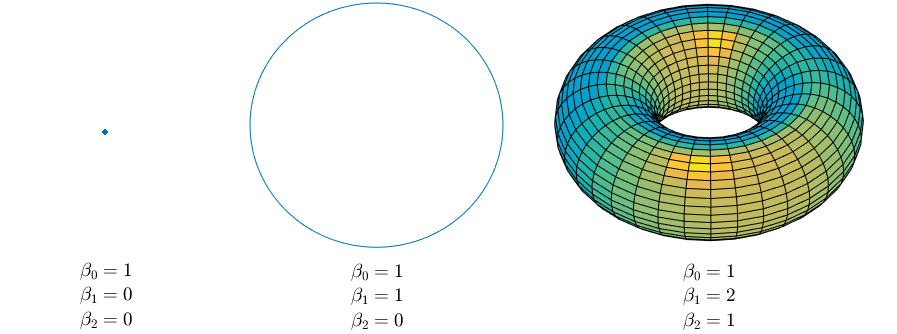} 
		\caption{Betti number representations of a point (left), circle (center), and torus (right).} 
		\label{fig:threebetti}
	\end{figure}

	Figure \ref{fig:threebetti} illustrates the Betti numbers of a point, circle, and torus (hollow donut). Each has a single connected component; the circle has a single loop; the torus has two loops (in the vertical and horizontal directions) and encloses a single solid void.  
	
	In order to perform computations on a topological space, one needs to discretize and represent the space with a finite amount of information.  Fortunately, this is possible for most spaces appearing in practice through the process of triangulation: we encode the space as a \emph{simplicial complex}, which means a collection of vertices, edges, triangles, tetrahedra, and higher-dimensional simplices, together with the data of how all these pieces are attached.  For instance, the circle is a single vertex with a single edge attached at both ends to the vertex; if we add a diagonal edge to each rectangular region of the torus depicted in Figure \ref{fig:threebetti} then we have decomposed the torus into a collection of attached triangles.  The computation of Betti numbers from a simplicial complex representation of a topological space is standard linear algebra. 

	\subsection{Persistent Homology}
	For a 1-parameter family of topological spaces, or simplicial complexes, \emph{persistent homology} offers a method of quantifying the dynamics of topological features (e.g., when holes appear and disappear).  A 1-parameter family of simplicial complexes where simplices are added but never removed is called a \emph{filtered} simplicial complex, and the parameter is then usually called a \emph{filtration} parameter.  

	A \emph{point cloud}---a finite, unordered collection of points in $\mathbb{R}^n$ or some other metric space---is uninteresting as a topological space since it is discrete: $\beta_0$ is the number of points in the cloud and $\beta_k = 0$ for all $k \ge 1$.  However, TDA introduces methods of building a filtered simplicial complex from a point cloud, thereby reconstructing multi-scale topological features of the object or distribution from which the points are sampled.  The standard approach is the \emph{Rips} complex: points in the cloud serve as vertices; pairs of points are joined by an edge when their distance is less than the filtration value; all possible higher-dimensional simplices are filled in (that is, whenever three points are connected by all three possible edges, a triangle is filled in, and whenever four points are joined by all six possible edges, a tetrahedron is filled in, etc.).
	
	Persistent homology can be visually summarized with the \emph{barcode plot}: for each dimension $k$, plot a collection of horizontal intervals whose left endpoint is the filtration value at which a particular $k$-dimensional homology class is born and whose right endpoint indicates its death.  The number of intervals over a filtration value is the Betti number $\beta_k$ at that value.  A more compact visual summary is the \emph{persistence diagram}: here each homology class is plotted as a point whose $x$-coordinate is the birth time and $y$-coordinate is the death time; different symbols are used to distinguish homological dimension.  The prominence, or persistence, of a topological feature corresponds to the length of a bar in a barcode plot, or to the height above the diagonal line $y=x$ in a persistence diagram. 
	 	
	\begin{figure}[!ht]
		\centering
		\begin{subfigure}[b]{0.4\textwidth}
			\centering
			\includegraphics[width=1\textwidth]{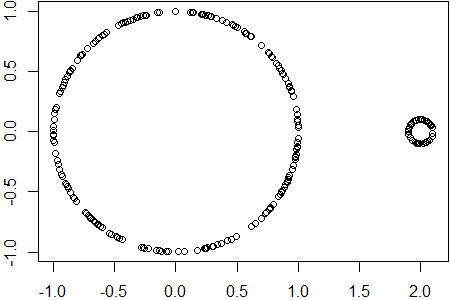}
			\caption{Scatterplot}\label{fig:exscatter1}
		\end{subfigure}
		\begin{subfigure}[b]{0.29\textwidth}
			\centering
			\includegraphics[width=1\textwidth]{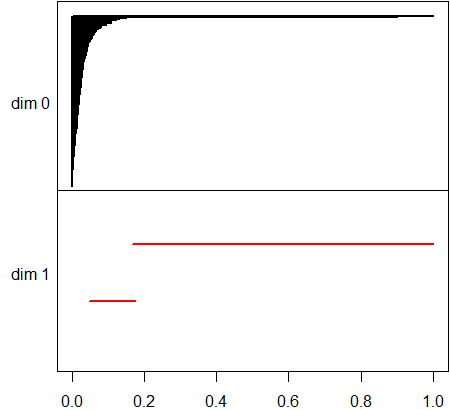}
			\caption{Barcode plot}\label{fig:exbp1}
		\end{subfigure}
		\begin{subfigure}[b]{0.29\textwidth}
			\centering
			\includegraphics[width=1\textwidth]{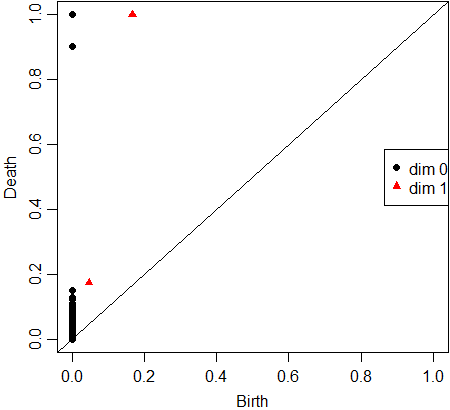}
			\caption{Persistence diagram}\label{fig:expd1}
		\end{subfigure}

		\begin{subfigure}[b]{0.4\textwidth}
			\centering
			\includegraphics[width=1\textwidth]{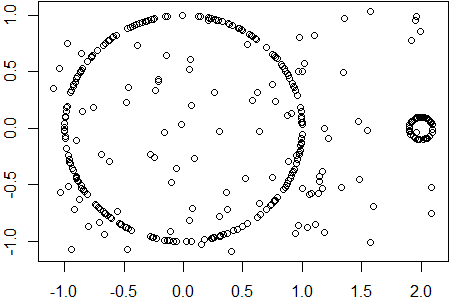}
			\caption{Scatterplot}\label{fig:exscatter2}
		\end{subfigure}
		\begin{subfigure}[b]{0.29\textwidth}
			\centering
			\includegraphics[width=1\textwidth]{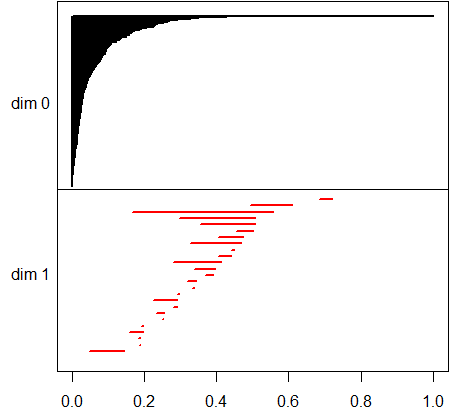}
			\caption{Barcode plot}\label{fig:exbd2}
		\end{subfigure}
		\begin{subfigure}[b]{0.29\textwidth}
			\centering
			\includegraphics[width=1\textwidth]{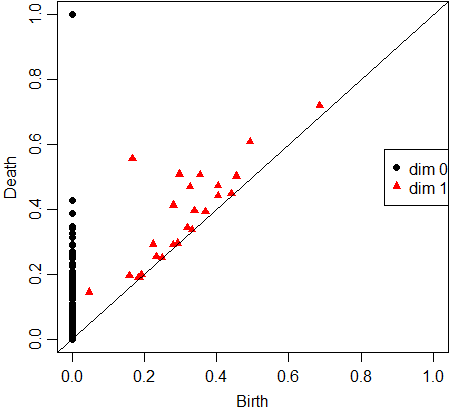}
			\caption{Persistence diagram}\label{fig:expd2}
		\end{subfigure}
		\caption{Scatterplots of points concentrating around two circles and the corresponding barcode plots and persistence diagrams.}\label{fig:ex}
	\end{figure}
	
	Figure~\ref{fig:ex} shows scatterplots, barcode plots and persistence diagrams obtained using the Rips complex.  Note that when there is no noise, the persistence of a dimension one feature (triangle plotted above the diagonal) is heavily affected by the radius of the circle it comes from; when noise is added, many false loops appear in the persistence diagram and it becomes difficult to infer the number of actual circles in the point cloud.

	\subsection{Robust Morse-Based Approach}
    To overcome sensitivity to noise and outliers, robust TDA approaches have been developed using distance to a measure \citep{Chazal2011,Chazal2014}, kernel distances \citep{Phillips2013}, kernel density estimators \citep{Fasy2014}, and kernel estimations \citep{Bobrowski2017}. These methods first transform a discrete point cloud into a continuous \emph{manifold} via a smoothing function.  For example, graphing the $m$th nearest neighbor function turns a point cloud in $\mathbb{R}^2$ into a surface in $\mathbb{R}^3$; the larger the value of $m$, the more rounded this surface will be.  The manifold thus obtained can be filtered by its super-level sets, sets above a threshold value.  The filtered super-level sets yield a 1-parameter family of topological spaces---which, when triangulated for computational purposes, is a filtered simplicial complex called the \emph{Morse} complex of the point cloud---so once again persistent homology can be computed.  The Morse complex is constructed with the smoothing function; the resulting persistent homology is more robust than that of the Rips complex.  In the Morse-based approach, choosing the appropriate smoothing parameter is important for topological inference.  \cite{Chazal2014} suggest a method for choosing the optimal smoothing parameter using information measures. 
	
		\begin{figure}[!ht]
		\centering
		\begin{subfigure}[b]{0.45\textwidth}
			\centering
			\includegraphics[width=0.65\textwidth]{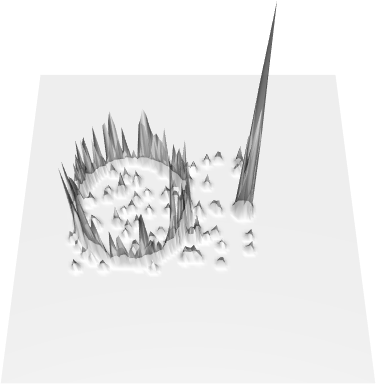}
			\caption{Manifold at bandwidth 0.03}\label{fig:exmorseman1}
		\end{subfigure}
		\centering
		\begin{subfigure}[b]{0.45\textwidth}
			\centering
			\includegraphics[width=0.67\textwidth]{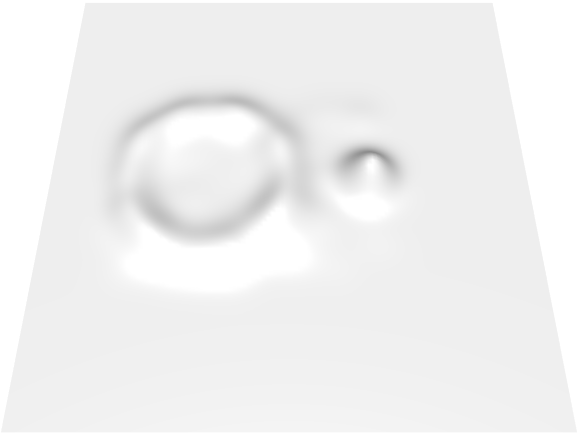}
			\caption{Manifold at bandwidth 0.2}\label{fig:exmorseman2}
		\end{subfigure}
		\centering
		\begin{subfigure}[b]{0.45\textwidth}
			\centering
			\includegraphics[width=0.9\textwidth]{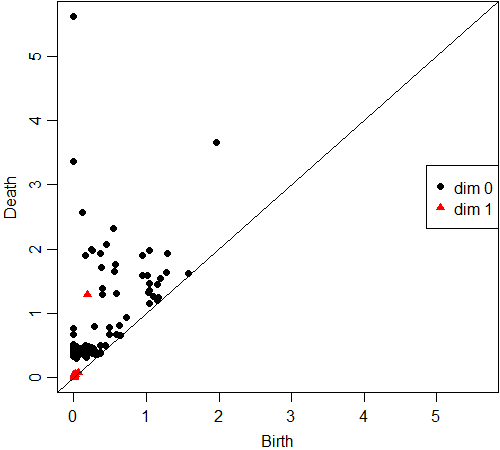}
			\caption{Persistence diagram at bandwidth 0.03}\label{fig:exmorse1}
		\end{subfigure}
		\centering  
		\begin{subfigure}[b]{0.45\textwidth}
			\centering
			\includegraphics[width=0.9\textwidth]{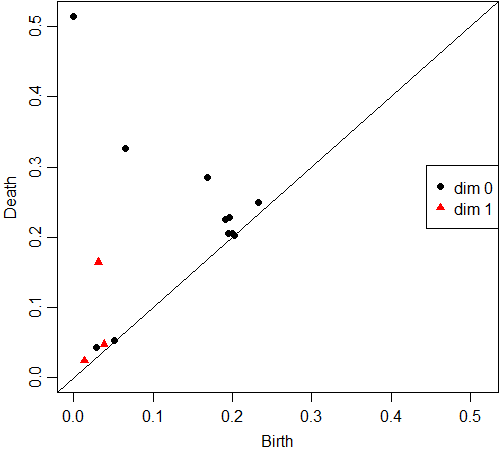}
			\caption{Persistence diagram at bandwidth 0.2}\label{fig:exmorse2}
		\end{subfigure}
		\caption{Manifolds and persistence diagrams of Morse filtration at two different smoothing parameters of two noisy circles data of Figure~\ref{fig:exscatter2}.}
		\label{fig:exmorse}
	\end{figure}
	
	There are disadvantages to these robust Morse-based smoothing approaches. First, there might not exist a single optimal smoothing parameter such that the corresponding persistent homology reveals features occurring at different scales.   For example, in Figure~\ref{fig:exmorse} we use the ``two noisy circles'' data from Figure~\ref{fig:exscatter2} and apply the kernel density estimator with two different bandwidth values.  From the persistence diagrams, we see that the noise has been cleaned up by switching from the Rips to the Morse complex---but for both smoothing parameter values, one of the two circles has been washed away in the process.  
		
	\begin{figure}[!ht]
		\centering
		\begin{subfigure}[b]{0.49\textwidth}
			\centering
			\includegraphics[width=0.9\textwidth]{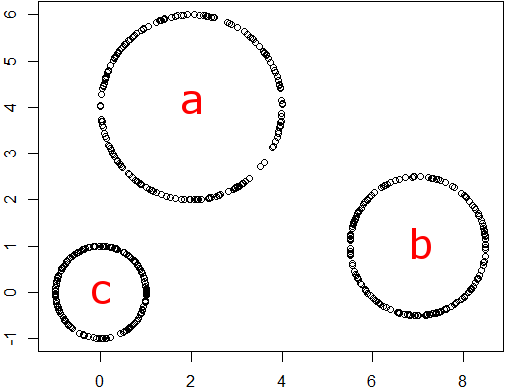}
			\caption{Scatterplot}\label{fig:exscatter}
		\end{subfigure}
		\centering  
		\begin{subfigure}[b]{0.45\textwidth}
			\centering
			\includegraphics[width=0.9\textwidth]{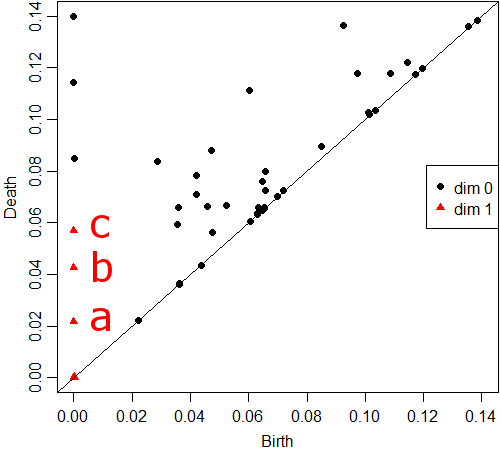}
			\caption{Persistence diagram of Morse filtration}\label{fig:expd}
		\end{subfigure}
		\caption{Scatterplot and its corresponding Morse persistence diagram at smoothing parameter 0.2 for three circles data. We label the three circles on the scatterplot and the corresponding $\beta_1$ elements on the persistence diagram}
		\label{fig:exsum}
	\end{figure}
	
	A second issue is that the Morse filtration only computes the number of $k$-dimensional holes of the level sets, not their sizes, so important scale information is lost.  Figure~\ref{fig:exsum} shows a point cloud with three circles, each containing 200 points, and its persistence diagram using kernel density estimator with bandwidth 0.2.  The Morse persistence diagram indicates three prominent loops in the data, but the height of each triangle above the diagonal line positively relates to the density of points on the corresponding circle and thus is negatively related to the radius.  The top triangle in the persistence diagram matches to the smallest circle, which has the highest point density, whereas the largest circle is represented as the bottom triangle. 
		
	\subsection{Significance of Features}
	The significance of a topological feature is generally recognized by the height-above-diagonal of a point in the persistence diagram.  Despite important steps toward a statistical theory of quantifying significance and producing confidence interval type analysis for persistent homology \citep{Fasy2014,Chazal2015}, serious conceptual challenges remain.  
	
	In a Rips complex, a small feature with high density is usually seen as insignificant because the Rips complex fills in the hole so quickly.  On the other hand, for the robust Morse-based approach, the significance depends, in addition to height-above-diagonal, on the smoothing parameter.  The smoothing parameter is chosen based on the size of data features one aims to uncover, but then height-above-diagonal in the persistence diagram reflects the density of points more than their significance.  Therefore, whether using Rips or Morse, it is impossible to fully capture significance with a single diagram: the significance of a feature depends on both its \emph{size} and \emph{density}. In the following section, we introduce the persistence terrace which is robust to noise and simultaneously reveals significance with regard to both size and point density of each topological feature.

	\section{Persistence Terrace}
	\label{sec:morse}
	The persistence terrace uses robust Morse-based persistent homology but incorporates a wide range of smoothing parameters instead of a single optimal value, similar to a scale space analysis \citep{Chaudhuri1999,Chaudhuri2000}.  
	For each dimension $k$, we plot a surface where the $x$-axis is the smoothing parameter, the $y$-axis is the filtration value, and the $z$-axis is the Betti number $\beta_k$.  Thus, for a point cloud in $\mathbb{R}^n$, a point on the persistence terrace with coordinates $(x_0,y_0,z_0)$ means there are $z_0$ holes of dimension $k$ on the Euclidean subset \[\{\vec{v}\in\mathbb{R}^n~|~f_{x_0}(\vec{v}) \ge y_0\},\] where $f_{x_0} : \mathbb{R}^n \rightarrow \mathbb{R}$ is the chosen smoothing function corresponding to parameter $x_0$.  In this paper, we use a Gaussian kernel density estimator as the smoothing function.  The smoothing parameter becomes the bandwidth.  Note that this subset is topologically equivalent to its graph in $\mathbb{R}^{n+1}$ under the function $f_{x_0}$, which is the super-level set used in Morse-based persistent homology. 
	
\begin{figure}[!ht]
		\centering
		\begin{subfigure}[b]{.45\textwidth}
			\includegraphics[width=1\textwidth]{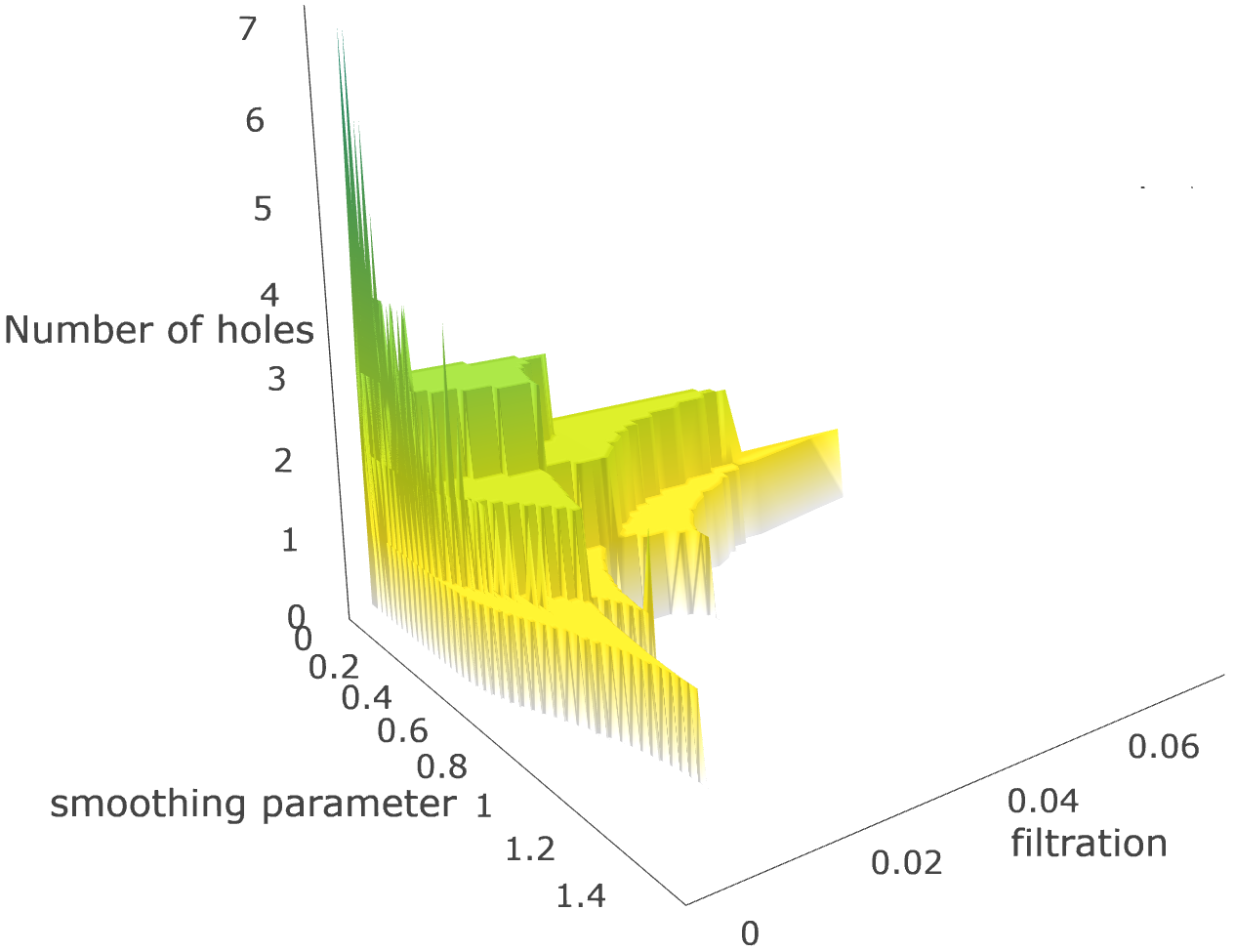}
			\caption{Overall view}\label{fig:ex3doverall}
		\end{subfigure}
		\begin{subfigure}[b]{.54\textwidth}
			\includegraphics[width=1\textwidth]{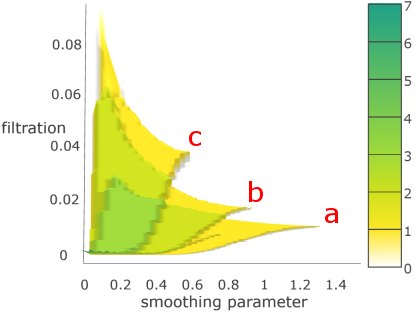}
			\caption{Satellite view}\label{fig:ex3dsatellite}
		\end{subfigure}
		\caption{$\beta_1$ persistence terrace for three circles data of Figure~\ref{fig:exscatter}. The labels of terrace layers in Figure~\ref{fig:ex3dsatellite} correspond to the circles in Figure~\ref{fig:exscatter}.}
		\label{fig:ex3d}
	\end{figure}
	
	Figure~\ref{fig:ex3d} shows the $\beta_1$ persistence terrace of the three circles point cloud from Figure~\ref{fig:exscatter} using 50 bandwidth values from 0.01 to 1.5.  Because Betti numbers are always integers, and they tend not to have gaps when the two parameters (smoothing and filtration) vary by small amounts, the surfaces plotted in a persistence terrace consist of flat layers that somewhat resemble a rice terrace (see Figure~\ref{fig:ex3doverall}).  Each $k$-dimensional topological feature in the point cloud contributes a layer to the persistence terrace; when there is a range of parameters for which multiple features are detectable, the corresponding layers in the terrace stack on top of each other and result in a higher altitude layer over their intersection.  
	
	The topological features in the point cloud are represented as terrace layers in the persistence terrace and the shape and location of the layers reflect the size and density of the corresponding features.  In the satellite view of a persistence terrace, the horizontal width ($x$-axis direction) is positively related to the size of the feature and the vertical length of a terrace layer ($y$-axis direction) is proportional to the point density.  Thus, we can match the topological features to the terrace regions according to their size and the point density.  For example, the large-sized but low density circle \emph{a} in Figure~\ref{fig:exscatter} is represented as the long horizontal width but short vertical length terrace region (layer \emph{a}) in Figure~\ref{fig:ex3dsatellite}.

	\begin{figure}[!ht]
		\centering
		\includegraphics[width=0.4\textwidth]{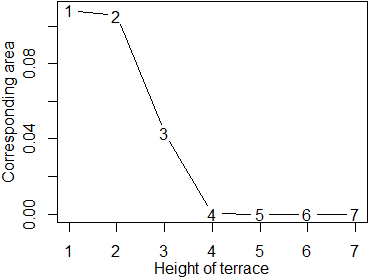}
		\caption{Terrace area plot of the persistence terrace in Figure~\ref{fig:ex3d}.}
		\label{fig:ex3ta}
	\end{figure}
	
	 When there is a small number of topological features, one can easily identify different layers.  However, when the number of features is large, it could be difficult to separate distinct terrace layers by eye.  We suggest a \emph{terrace area plot} to aid in determining the significant features.  The terrace area plot presents the area of terrace layers of a certain height, except the height-zero layer, in the persistence terrace.  Because the scales of filtration and smoothing parameter vary from dataset to dataset, we standardize each axis so that the total area is one.  Figure~\ref{fig:ex3ta} shows the areas of the terraces of the persistence terrace in Figure~\ref{fig:ex3d}.  The areas of the layers higher than three quickly settle down to zero.  Those small areas correspond to the spikes near the origin resulting from noise.  We can conclude that the layers of height one to three contribute to the significant features.  The terrace area plot can be used to select significant layers similar to a scree plot for a principal component analysis.  However, the areas in the terrace area plot can increase unlike the variances in the scree plot.    Note that the largest height among the selected layers, three in Figure~\ref{fig:ex3ta}, is the minimum number of significant features for which the corresponding terraces overlap each other.  If one can find layers that do not overlap with the selected height layer, then they can be counted as additional significant features.  

	\begin{figure}[!ht]
		\centering
		\includegraphics[width=0.9\textwidth]{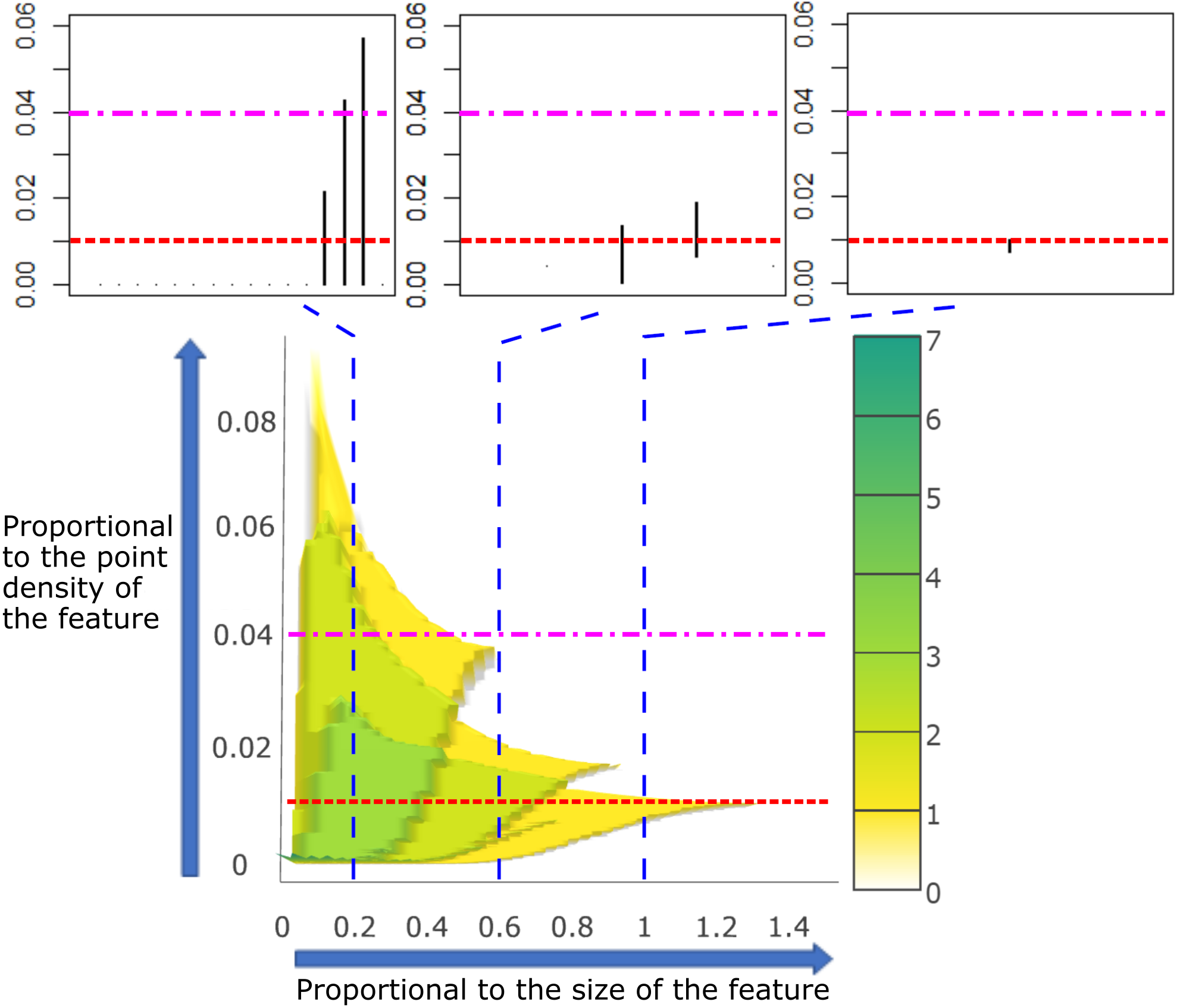}
		\caption{Barcode plots at smoothing parameters 0.2, 0.6 and 1, corresponding to vertical slices of the persistence terrace. Although barcode plots traditionally use the filtration value for the $x$-axis, we rotate the barcode counterclockwise 90 degrees to match up with the $y$-axis of the terrace.}
		\label{fig:ex3dbarcode}
	\end{figure}

	The relationship between the persistence terrace and barcodes is illustrated in Figure~\ref{fig:ex3dbarcode}.  Fixing a value of the smoothing parameter corresponds to taking a vertical slice of the persistence terrace.  While this slice of the terrace shows the Betti number $\beta_k$ at each filtration value, the barcode indicates this Betti number with $\beta_k$ separate horizontal intervals.  In Figure~\ref{fig:ex3dbarcode} we have chosen three different values of the smoothing parameter (0.2, 0.6 and 1) at which to slice the terrace and draw the corresponding barcode.

	\begin{figure}[!ht]
		\centering
		\begin{subfigure}[b]{.47\textwidth}
			\includegraphics[width=1\textwidth]{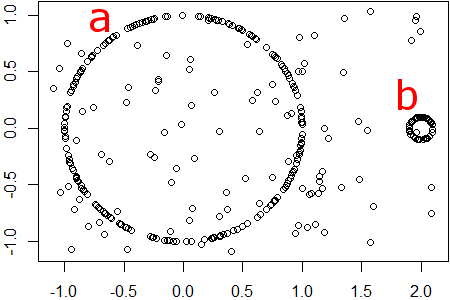}
			\caption{Scatterplot}\label{fig:exscatter22}
		\end{subfigure}
		\begin{subfigure}[b]{.51\textwidth}
			\includegraphics[width=1\textwidth]{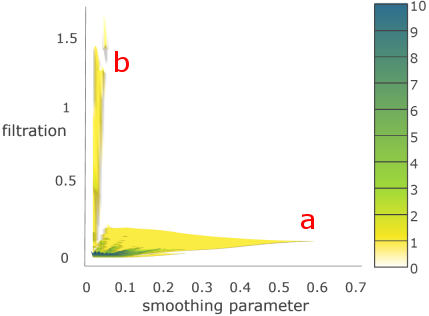}
			\caption{Persistence terrace (satellite view)}\label{fig:exospt}
		\end{subfigure}
		\caption{Scatterplot of the noise-added two circle data and $\beta_1$ persistence terrace.}
		\label{fig:exos}
	\end{figure}    
	
	The persistence terrace allows us to separate out topological features even when there is no value of the smoothing parameter that could detect all features.  Recall that in Figure~\ref{fig:exmorse} we tried two different values of the smoothing parameter, and in each case lost the information of one of the two circles; large-sized low density and small-sized high density circles.  This suggests that there is no single optimal smoothing parameter that reveals all features in the data.  Nonetheless, the persistence terrace in Figure~\ref{fig:exospt} shows two clear, distinct height-one layers (one vertically oriented, one horizontally oriented); any optimal range of smoothing parameter is so small it is essentially impossible to locate.  Therefore, one readily infers two distinct loops, one large and the other smaller and denser, in the data even though the standard Rips approach fails due to noise and the robust Morse approaches fail due to inadequacies of smoothing.


	\subsection{Computational algorithm}
	\label{subsec:comput}
	We provide the R package ``pterrace'' for the computation.  The computation of a persistence terrace can be divided into three algorithmic steps:
	\begin{equation*}
	\mbox{Point cloud data} \overset{Step\mbox{ }1} \longrightarrow \mbox{Barcodes} \overset{Step \mbox{ }2}\longrightarrow \mbox{Betti numbers} \overset{Step \mbox{ }3}\longrightarrow \mbox{Persistence terrace}
	\end{equation*}
	
	First, we compute the barcodes for the Morse complexes corresponding to a pre-defined vector of smoothing parameter values, using a specified smoothing function.  This is accomplished using a simple ``for loop'' and any of the pre-existing persistent homology software packages that includes Morse complex barcodes.  We use the R package TDA \citep{TDA} for this step.  Second (Algorithm 1 in Supplementary Material), we fix a dimension $k$ and for each fixed value of the smoothing parameter compute the Betti number $\beta_k$ using the $k$-dimensional barcodes.  Third (Algorithm 2 in Supplementary Material), we use the fact that for each fixed smoothing parameter value the function $\beta_k$ just computed is a step function, only changing at the filtration values computed in the previous step, in order to assemble all the Betti numbers into the persistence terrace.  
	

	\subsection{Detection of Features with Different Densities}
	\label{subsec:density}
	We can use the persistence terrace to identify differences in the densities of data points that make up the various topological features in the point cloud.  As discussed earlier, the persistence terrace can often be visually decomposed into height-one layers that overlap with each other in various regions.  The length of each height-one layer along the $y$-axis (i.e., the range of filtration values for which the corresponding topological feature persists) positively relates to the density of points in that feature.  Indeed, high density means that the smoothing function will take large values, so the manifold it produces will be very tall over high density regions and consequently that portion of its level set topology will remain constant for a large range of filtration values.
	
	\begin{figure}[!ht]
		\centering
		\begin{subfigure}[b]{0.49\textwidth}
			\centering
			\raisebox{-\height}{\includegraphics[width=0.9\textwidth]{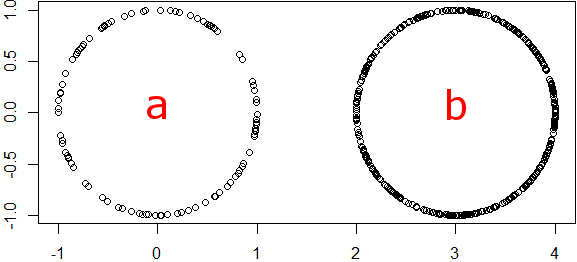}}
			\vspace{.6ex}
			\raisebox{-\height}{\includegraphics[width=1\textwidth]{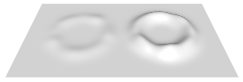}}
			\caption{Scatterplot (top) and manifold from bandwidth $0.2$ (bottom)}\label{fig:densityscatter}
		\end{subfigure}
		\centering  
		\begin{subfigure}[b]{.50\textwidth}
			\includegraphics[width=1\textwidth]{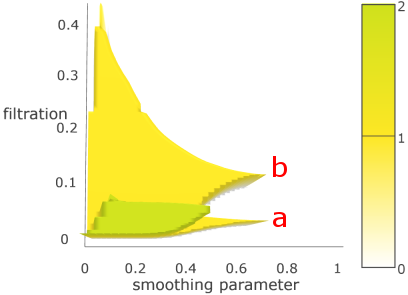}
			\caption{Satellite view of $\beta_1$ persistence terrace}\label{fig:density3dsatellite}
		\end{subfigure}
		\caption{Scatterplot, manifold and persistence terrace of two circles data with  different densities. The high density circle \emph{b} smooths to a high altitude volcano and appears as a persistence terrace layer stretching along the $y$-axis in the satellite view (layer \emph{b}).}\label{fig:densitysum}
	\end{figure}	
	
    To illustrate this, we generate 100 and 400 random points from two equal sized circles. Figure~\ref{fig:densityscatter} shows how the higher point density of the second circle is manifest as a higher altitude volcano-shaped manifold after smoothing with bandwidth $0.2$.  In the $\beta_1$ persistence terrace, we see two height-one terrace layers with a clear region of height-two overlap.  Looking at the satellite view of the persistence terrace in Figure~\ref{fig:density3dsatellite}, we see that both layers reach horizontally to equal maximal values of the smoothing parameter (which, as we discuss in the next subsection, stems from the fact that the circles have the same radius), but one layer reaches much further vertically along the filtration axis.  This taller terrace layer represents the denser circle since for large values of the filtration parameter the super-level set of the tall volcano will be an annulus (which, topologically, is a circle) while that of the short volcano will be empty.  If we were to remove the dense circle from the point cloud and re-plot the persistence terrace, we would see this tall terrace removed (so the height-two overlap would drop down to height one) and the shorter terrace layer would remain unchanged.  Thus the two topological signals have been completely disentangled and identified by their corresponding point densities.

	\subsection{Detection of Features with Different Sizes}
	\label{subsec:size}	
	We can also infer the size of topological features in a point cloud from the persistence terrace. Just as density was measured by how far a terrace layer stretches along the $y$-axis in the satellite view, size is measured by how far it stretches along the $x$-axis, by the range of smoothing parameters for which the corresponding feature is detectable.  Indeed, as the smoothing parameter increases, the corresponding manifolds become more rounded so the finer topological features of the super-level sets get washed away.  
	
		\begin{figure}[!ht]
			\centering
			\begin{subfigure}[b]{0.49\textwidth}
				\centering
				\raisebox{-\height}{\includegraphics[width=0.7\textwidth]{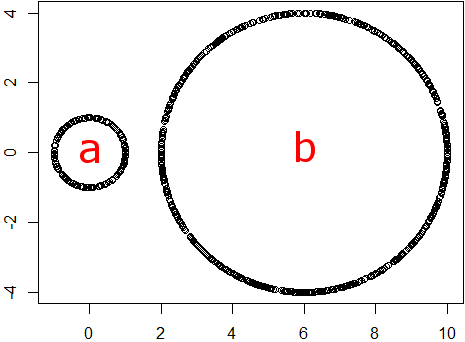}}
				\raisebox{-\height}{\includegraphics[width=0.85\textwidth]{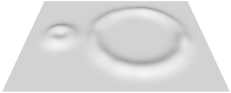}}
				\caption{Scatterplot (top) and manifold from bandwidth $0.5$ (bottom)}\label{fig:sizescatter}
			\end{subfigure}
			\centering  
			\begin{subfigure}[b]{.5\textwidth}
				\includegraphics[width=1\textwidth]{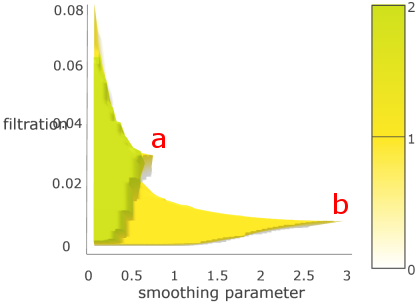}
				\caption{Satellite view of $\beta_1$ persistence terrace}\label{fig:size3dsatellite}
			\end{subfigure}
			\caption{Scatterplot, manifold and persistence terrace of two circles with equal density but unequal radius data.  The volcano manifold from the small circle (circle \emph{a}) fills in more quickly as the smoothing parameter is increased and appears as a persistence terrace layer stretching along a narrow stretch of the $x$-axis in the satellite view (layer \emph{a}).}\label{fig:sizesum}
		\end{figure}
	
	To illustrate this, we generate 200 points on a radius 1 circle and 800 points on a radius 4 circle so that the two circles have almost the same point densities.  We see in Figure \ref{fig:sizescatter} that they smooth into volcanoes of similar height.  The $\beta_1$ persistence terrace in Figure~\ref{fig:size3dsatellite} clearly consists of two overlapping height-one layers.  These layers reach to roughly comparable heights along the $y$-axis since the two densities agree, but one terrace layer reaches much further along the $x$-axis: the volcano coming from the smaller circle fills in more quickly than that of the larger circle as the smoothing parameter is increased.  
	

  	\section{Simulation Study}
	\label{sec:sim}
	
	The data sets in the previous section are deliberately very simple in order to focus the discussion and analysis of persistence terraces on specific attributes and behaviors.  In this section we explore two data sets that are less artificial: one involves different shapes and more noise/distortions, the other comes from medical imaging.

	\subsection{Features with Noise}   
\begin{figure}[!ht]
		\centering
		\begin{subfigure}[b]{0.49\textwidth}
			\centering
			\includegraphics[width=0.9\textwidth]{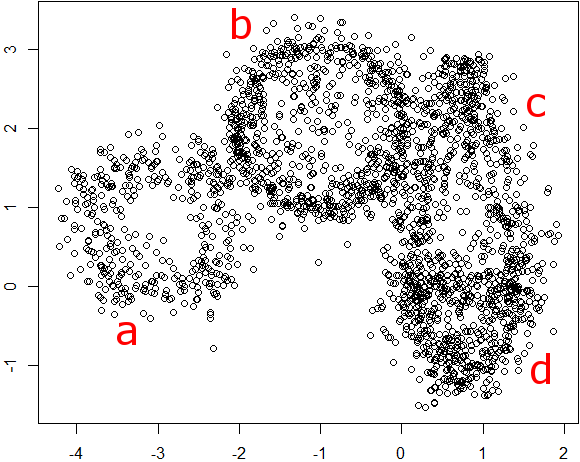}
			\caption{Scatterplot}\label{fig:noisescatter}
		\end{subfigure}
		\centering  
		\begin{subfigure}[b]{0.49\textwidth}
			\centering
			\includegraphics[width=0.85\textwidth]{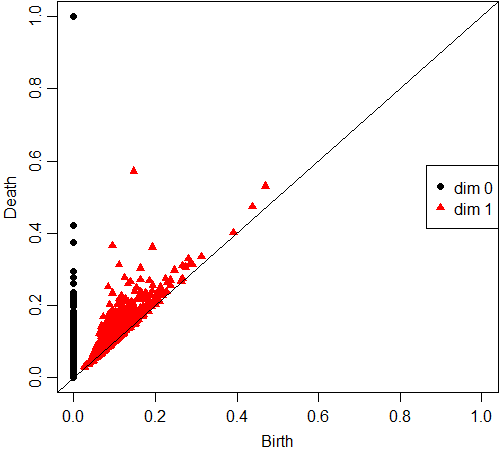}
			\caption{Persistence diagram}\label{fig:noisepd}
		\end{subfigure}
		\caption{Scatterplot of four noisy shapes data and the Rips persistence diagram.}\label{fig:noisesum}
	\end{figure}

	Here we demonstrate that the persistence terrace can be used to help analyze data that are both noisy and less uniform than in the previous section.  We generate a planar point cloud with points clustering around four ``holes'' so that four topological loops should be present (see Figure~\ref{fig:noisescatter}).  Specifically, we place 400 points uniformly randomly on a $1\times 1$ square then perturb these points with random $N(0,0.15)$ noise (square \emph{a}). We create 800 points around a radius 1 circle \emph{b}: for inside and outside of the circle, 400 points are generated that follow an exponential distribution with rates 4 and 10, respectively.  We also create an equilateral triangle \emph{c} and an isosceles triangle \emph{d}, by placing 200 points randomly uniformly on each triangle edge, but then perturb these points with $N(0,0.15)$ noise. The persistence diagram in Figure~\ref{fig:noisepd} is computed using the Rips complex.  While some triangles in that diagram are far above the diagonal line, suggesting persistent loops in the data, there is far too much noise to be able to infer the correct number of loops, namely $\beta_1 = 4$.

\begin{figure}[!ht]
		\centering
		\begin{subfigure}[b]{0.4\textwidth}
			\centering
			\includegraphics[width=1\textwidth]{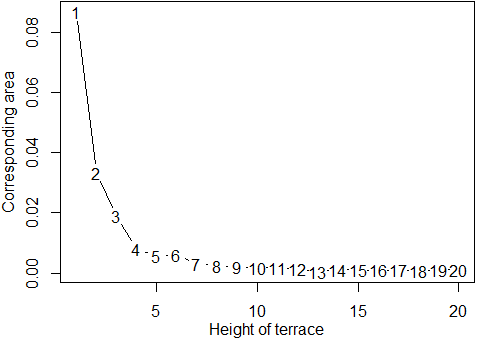}
			\caption{Terrace area plot}\label{fig:noiseta}
		\end{subfigure}
		\begin{subfigure}[b]{0.54\textwidth}
			\centering
			\includegraphics[width=1\textwidth]{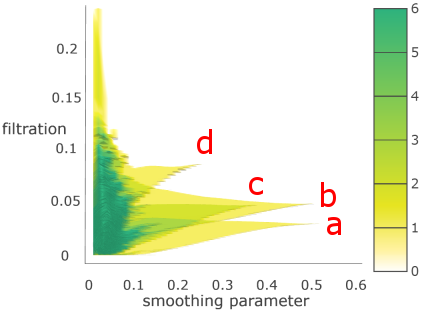}
			\caption{$\beta_1$ persistence terrace}\label{fig:noisept}
		\end{subfigure}
		
		\caption{Terrace ara plot and satellite view of $\beta_1$ persistence terrace for the noisy four shape data.  There are four prominent layers, corresponding to the four shapes, though one is stacked on another.  The full version of the terrace area plot is drawn in Figure 1 in Supplementary Material.}
		\label{fig:noise3d}
	\end{figure}

	We compute the $\beta_1$ persistence terrace using 50 smoothing parameter values evenly distributed between 0.01 to 0.6.  The persistence terrace detects up to 83 loops in the data, though most of these only occur for relatively small values of the smoothing parameter.  We show the terrace area plot up to height 20 and persistence terrace in Figure~\ref{fig:noise3d}.  The terrace area plot in Figure~\ref{fig:noiseta} implies that the areas of layers of height one to three are larger than the other heights.  Thus we can conclude that there are at least three significant features in the data.  Although we expect the layers of height greater than three to be considered noise, we put all levels greater than six into a single category in the persistence terrace Figure~\ref{fig:noisept} to avoid erroneous interpretation.  The height-three layer in the persistence terrace is an overlap of the layers \emph{a}, \emph{b}, and \emph{c}.  The layer \emph{d} is a distinct layer that is apart from the height-three region.  Therefore, we can consider the layer \emph{d} as a representation of an additional significant feature, and that makes the total number of features in the data four. 
	
    In addition, we can use the methods from the previous section to give a rough, qualitative description of the four loops identified by the persistence terrace.  The two largest loops (layer \emph{a} and \emph{b}) have roughly the same size though one has slightly greater point density; this latter density matches the density of another, slightly smaller loop (layer \emph{c}); finally, the fourth loop (layer \emph{d}) is both smaller and denser than the rest.  This description appears to accord with the square, circle, isosceles triangle and equilateral triangle, respectively.

	\subsection{Counting Muscle Fibers}
	\label{sec:mf}
    Muscle tissue consists of tube-like shapes known as muscle fibers which are bundles of filaments ensheathed by a connective tissue known as endomysium.  A cross-section of a muscle thus reveals a collection of semi-homogeneous regions (if one blurs the filaments together), one for each muscle fiber, that are delineated by walls made of endomysium.  Counting the number of muscle fibers in a cross-sectional slice of a tissue sample can, therefore, be viewed as a topological problem: we need to compute the number of independent loops formed by the endomysium.  Since the sizes of the loops vary and the cross-sectional image is bound to have noise, this is a natural setting to apply the persistence terrace.  For this example, we adapt Figure 1 of \cite{Mula2013}, the muscle tissue cross-section image.  

	\begin{figure}[!ht]
		\centering
		\includegraphics[width=0.45\textwidth]{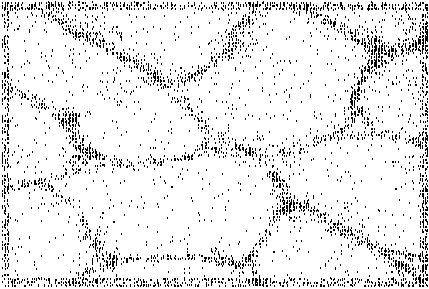}  
		\caption{Scatterplot of 6,500 points sampled from the muscle tissue cross-sectional image.}
		\label{fig:mfscatter}
	\end{figure}

	The first step in applying the persistence terrace method is to convert the cross-sectional image into a point cloud.  Since we are interested in the loops made from the endomysium, we should sample points from the endomysium in the cross-sectional image.  To do this, we first convert the original RGB image to a grayscale intensity image.  Next, we randomly select 5,000 endomysial points from the grayscale image.  We sample in proportion to the grayscale intensity so that darker pixels have a higher likelihood to be selected.  Since the cross-sectional image contains several incomplete muscle fibers at the boundary that we would like to include in the count, we add lines around the boundary of the black-and-white image to artificially close off all the broken loops.  For the particular cross-sectional image we are using here, this results in 11 muscle fibers, including these partial boundary fibers.  We generate an additional 1,500 points by randomly sampling from the boundary.  The resulting point cloud is shown in Figure~\ref{fig:mfscatter}.  

	Note that the muscle fiber cross-sections vary considerably in size, shape, and convexity.  Note also that there are two types of noise: black pixels within the muscle fibers and white pixels within the endomysium, both due to sampling error.  There is a further, more substantial complication to the data: while muscle fibers are packed together rather tightly, there are nonetheless small gaps where multiple fibers come together.  Biologically, this means there are small chambers enclosed by endomysium that are devoid of filaments and thus not considered muscle fibers.  From a data perspective, this means there are small white regions within the walls that lead to small loops in the point cloud that should not contribute to the muscle fiber count.  The speckling noise renders Rips persistent homology inadequate while these gap loops make it nearly impossible to choose a single optimal smoothing parameter.   These are both motivations for using the persistence terrace.
	
	\begin{figure}[!ht]
		\centering
		\begin{subfigure}[b]{0.45\textwidth}
			\centering
			\includegraphics[width=1\textwidth]{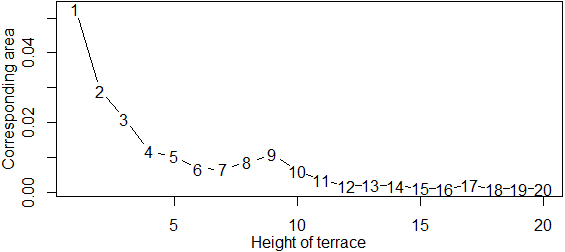}
			\caption{Terrace area plot}\label{fig:mfta}
		\end{subfigure}
		\begin{subfigure}[b]{0.6\textwidth}
			\centering
			\includegraphics[width=1\textwidth]{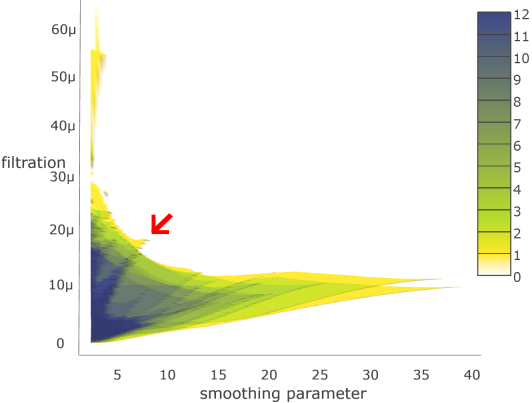}
			\caption{$\beta_1$ persistence terrace}\label{fig:mfpt}
		\end{subfigure}
		\caption{Terrace area plot and satellite view of the $\beta_1$ persistence terrace of the point cloud sampled from the muscle tissue cross-sectional image.  The full version of the terrace area plot is given in Figure 2 in Supplementary Material}\label{fig:mfl}
	\end{figure}
	
	To build the $\beta_1$ persistence terrace, we use 100 smoothing parameter values between 2 and 40.  Figure~\ref{fig:mfl} shows the terrace area plot and persistence terrace.  The terrace area plot shows that the minimum number of significant loop is either 9 or 10; the areas greater than 9 or 10 are small enough to be considered as noise.  We color the heights greater than 12, considered to be noise, as a single category in the persistence terrace Figure~\ref{fig:mfpt}.  Also, we can find a height-one triangular region, appears around filtration value 20$\mu$ and smoothing parameter 8 (indicated by an arrow), which does not overlap with the height 9-10 terrace region.  This suggests that we can count one additional small-sized high point density muscle fiber.  Therefore, by considering the additional triangular region, we find 10-11 muscle fibers in total. 
	
	
	
	We can also see that there is a fairly prominent vertically oriented region extending to filtration value $60\mu$, giving terrace heights in the range 1-3. This implies that there is a small sized but high point density loop.  The terrace region may correspond to the muscle fiber gaps (loops in the endomysium that do not enclose any filaments).  It is difficult to get a precise reading of all the gaps in the scatterplot of Figure~\ref{fig:mfscatter} because they are filled with noise pixels.  These non-fiber loops are small in size but have comparable point density to the actual muscle fibers.  Thus, they correspond to the top-left region in the persistence terrace.  So while this portion of the analysis is more murky, it does seem that the persistence terrace is also able to detect and count some of the muscle fiber gaps.	
	
	\section{Discussion}
	\label{sec:disc}
	
	\subsection{Inference Tool for Persistence Terrace}
	
	In the persistence terrace, topological features in point cloud data are represented as terrace layers and we can make a robust estimation of the feature including its size and point density.  
	However, when there is a large number of features, it is difficult to disentangle the individual layers.  We suggest the terrace area plot be used as an aid for determining the number of significant features, but one still needs to pay attention to the terrace layers to make a correct interpretation.  In a future paper, we plan to develop a method to label the different layers systematically, which can help analyze data with numerous topological features.

	\subsection{Algorithmic Improvements}
	To create a persistence terrace, we compute the Morse complex persistent homology for each value of the smoothing parameter specified in a pre-determined range.  For large data sets, each persistent homology computation can be intensive. 
	Fortunately, since these persistent homology computations are all independent, the procedure is embarrassingly parallelizable: just compute persistent homology for each smoothing value on separate processors then assemble.  In our R package, we provide the parallel computing option. 

	We typically choose the range of smoothing parameter values heuristically based on a rough estimate of the features in the point cloud one hopes to uncover, and then subdivide this range into equally sized intervals.  We recommend to start from using the large range of smoothing parameters and then narrow down the range that features appear on the persistence terrace. 
	
    Also, a subtle question is how to choose the number of intervals: too many intervals and the computational time becomes unreasonable, too few and the persistence terrace looks coarse and becomes difficult to read.  Figure 3 in Supplementary Material shows four persistence terraces computed from the Figure~\ref{fig:exscatter} data, using 25, 50, 75, and 100 smoothing parameter values.  When the number of features in data is relatively small, then a small number of intervals is enough.  On the other hand, when data are noisy and include a large number of features as Figure~\ref{fig:mfscatter} data, then the larger number of intervals is required.
	
	\subsection{Varying the Dimensions}
	In this paper we have focused on the $\beta_1$ persistence terraces for planar point clouds.  Extending from point clouds in $\mathbb{R}^2$ to arbitrary $\mathbb{R}^n$ is trivial: all steps in the algorithms already allow for this possibility, as does the qualitative analysis of the resulting persistence terraces.  
	One can also extend from $\beta_1$ topological features (i.e., loops) to any other dimension.  For higher dimensions, 
	the overall analysis should be quite similar.  Most applications of TDA in the literature focus on $\beta_1$ and $\beta_0$, so we did not explore $\beta_2$ or higher persistence terraces in this paper but one could certainly find applications.  On the other hand, we tried some experiments with $\beta_0$ persistence terraces and were unable to draw any conclusions.  Since $\beta_0$ is the number of connected components, the $\beta_0$ persistence terraces should encode multi-scale clustering information about the point cloud, but it is limited by the fact that clustering involves more than just point density.  We will explore this point in future work.

	\begin{center}
		{\large\bf SUPPLEMENTARY MATERIAL}
	\end{center}
	
	\begin{description}
		
		\item[Title:] Algorithms and Figures (pdf)
		
		\item[R-package for persistence terrace:] R-package ``pterrace'' containing code to plot the persistence terrace. The package also contains all datasets used as examples in the article. (GNU zipped tar file)
		
	\end{description}
	
	\bibliographystyle{Chicago}
	\bibliography{Persistence_Terrace_JCGS_References}

\end{document}


\begin{center}
		{\Large Supplementary Material}
	\end{center}

	\begin{algorithm}[!ht]
		\begin{algorithmic}
			\STATE \textbf{Input} $sp$ (smoothing parameter vector) and $barcodes$ (sets of barcodes)
			\STATE $kholes \leftarrow$ NULL (List of Betti number values and locations, for each smoothing parameter value)
			\FOR{$i = 1 \rightarrow $ length($sp$)}
			\STATE $kbarcode \leftarrow barcodes[[i]] \{ \mbox{ dim}=k,\mbox{ birth, death }\}$ (Select $k$-dimensional barcode obtained from $i$th smoothing parameter value)
			\STATE $m \leftarrow \mbox{column length of kbarcode}$ (Number of $k$-dimensional bars)
			\IF{$m=0$ (No $k$-dimensional bars) }
			\STATE $track\leftarrow \{$ filtration=0, numk$=0\}$ 
			\ELSE
			\STATE filtration $\leftarrow \{$ birth values in $kbarcode$; death values in $kbarcode \mbox{ } \}$ 
			\STATE kBetti $\leftarrow \{ \mbox{ } \textbf{1}_{\textbf{m}};\textbf{-1}_{\textbf{m}} \mbox{ } \}$  			\STATE $track\leftarrow \{$ filtration, kBetti $\}$ ( Bind two vectors into $(2m \times 2)$ matrix, thereby assigning $1$ and $-1$ to the birth and death points, respectively)
			\STATE Sort $track$ matrix by `filtration' in ascending order
			\STATE $track \$ \mbox{kBetti}  \leftarrow \mbox{cumsum}(track \$ \mbox{kBetti} ) $ (Replace the kBetti column by its cumulative sum)
			\ENDIF
			\STATE $kholes[[i]]\leftarrow track$
			\ENDFOR
			\RETURN $kholes$
		\end{algorithmic}
		\caption{Compute Betti number values and locations from barcodes}
		\label{alg2}
	\end{algorithm}

\newpage
	
	\begin{algorithm}[!ht]
		\begin{algorithmic}
			\STATE \textbf{Input} $sp$ (smoothing parameter vector) and $kholes=\{$ filtration, $k$th Betti number $\}$
			\STATE $n \leftarrow$ length($sp$)
			\STATE $xvec \leftarrow$ sp ($x$ values vector: smoothing parameters)
			\STATE $yvec \leftarrow$ NULL ($y$ values vector: filtration values when the $k$th Betti number change)

			\FOR{$i=1 \rightarrow n$}
			\STATE $yvec \leftarrow \{ \mbox{ } yvec,\mbox{ } kholes[[i]] \$ \mbox{filtration}  \mbox{ } \} $ (Stack all filtration values)
			\ENDFOR
			\STATE $yvec \leftarrow$ sort($yvec$) (Sort $yvec$ in descending order)
			
			\STATE $zmat \leftarrow$ $\textbf{0}_{\textbf{length(xvec)} \times \textbf{length(yvec) }}$ ($z$ values matrix: $k$th Betti number)            
			
			\FOR{$p = 1 \rightarrow n$}
			\STATE $filtration \leftarrow kholes[[p]]\$filtration$
			\STATE $kBetti \leftarrow kholes[[p]]\$kBetti$
			\STATE $zvec \leftarrow \mathbf{0_{length(yvec)}}$
			\FOR{$q = 1 \rightarrow $ length$(kBetti)$}
			\STATE $zvec=zvec+(filtration[q+1] < yvec)*(yvec \leq filtration[q])*kBetti[q]$ (Fill out $k$th Betti numbers for all filtration values)
			\ENDFOR
			\STATE $zmat[,p] \leftarrow zvec$ (Save $zvec$ to $p$th column of $zmat$)
			\ENDFOR
			\RETURN $[xvec,yvec,zmat]$ 
		\end{algorithmic}
		\caption{Compute persistence terrace matrix from Betti number values/locations} 
		\label{alg3}
	\end{algorithm}

\newpage


 \begin{figure}[!ht]
	\centering
	\includegraphics[width=0.9\textwidth]{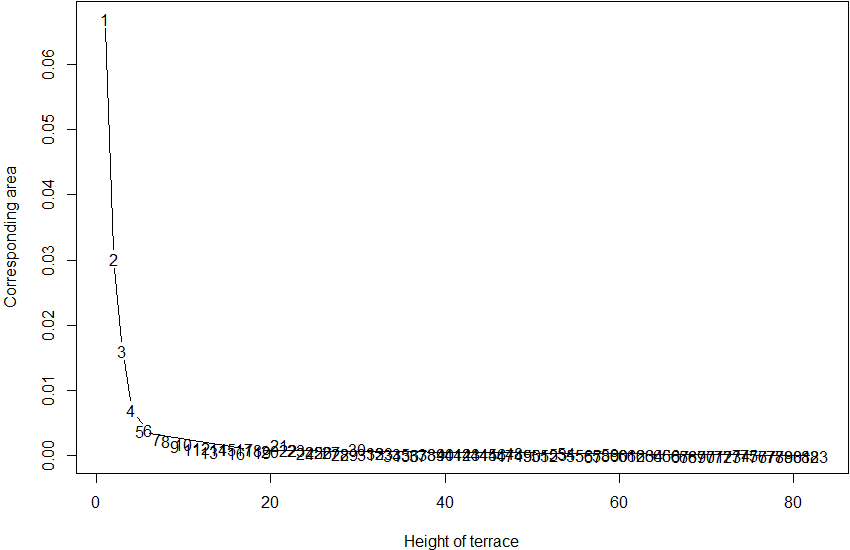}
	\caption{Terrace area plot for the simulated data set with four noisy features.}\label{fig:nfta}
\end{figure}

 \begin{figure}[!ht]
	\centering
	\includegraphics[width=0.9\textwidth]{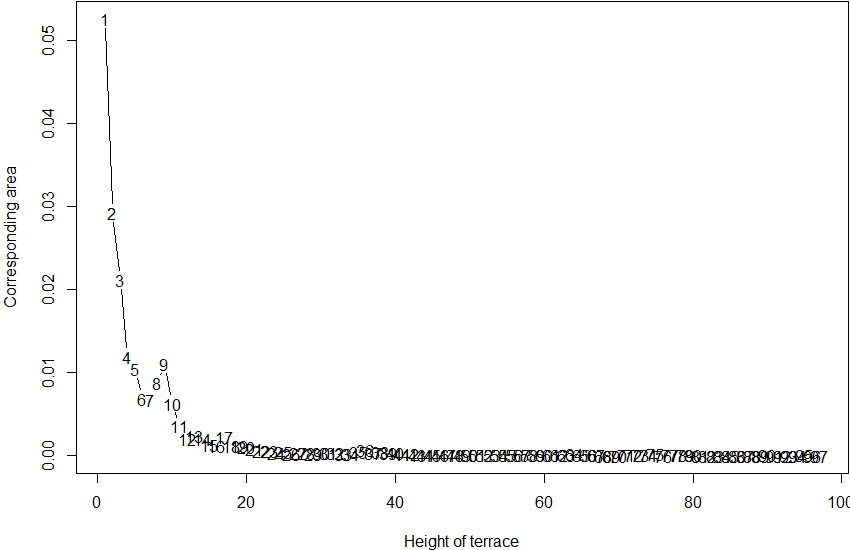}
	\caption{Terrace area plot for the muscle fiber data}\label{fig:mfta}
\end{figure}

 \begin{figure}[!ht]
	\centering
	\begin{subfigure}[b]{0.49\textwidth}
		\centering
		\includegraphics[width=1\textwidth]{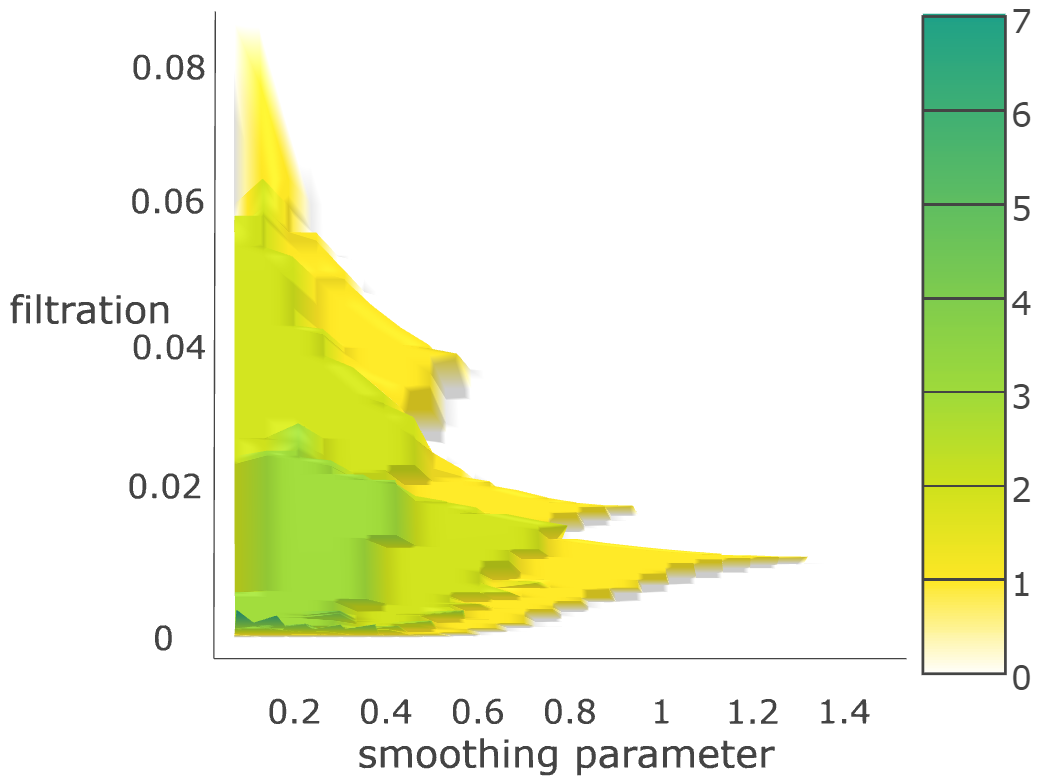}
		\caption{25 smoothing parameters}\label{fig:pt25}
	\end{subfigure}
	\centering  
	\begin{subfigure}[b]{0.49\textwidth}
		\centering
		\includegraphics[width=1\textwidth]{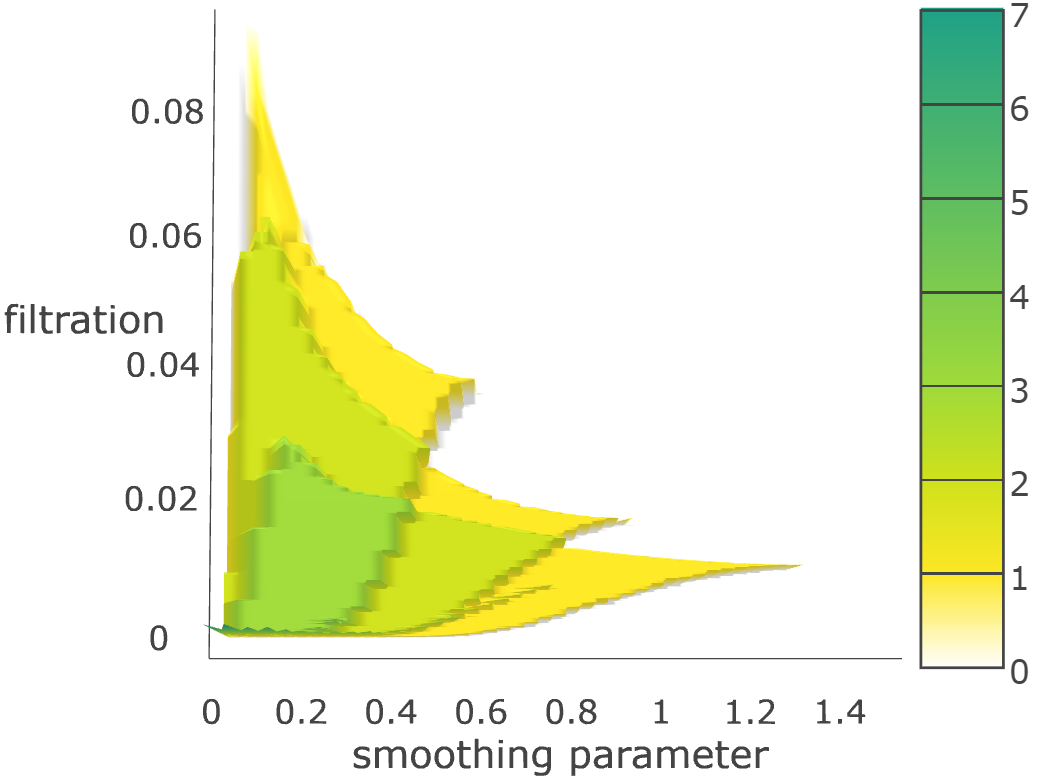}
		\caption{50 smoothing parameters}\label{fig:pt50}
	\end{subfigure}
	\begin{subfigure}[b]{0.49\textwidth}
		\centering
		\includegraphics[width=1\textwidth]{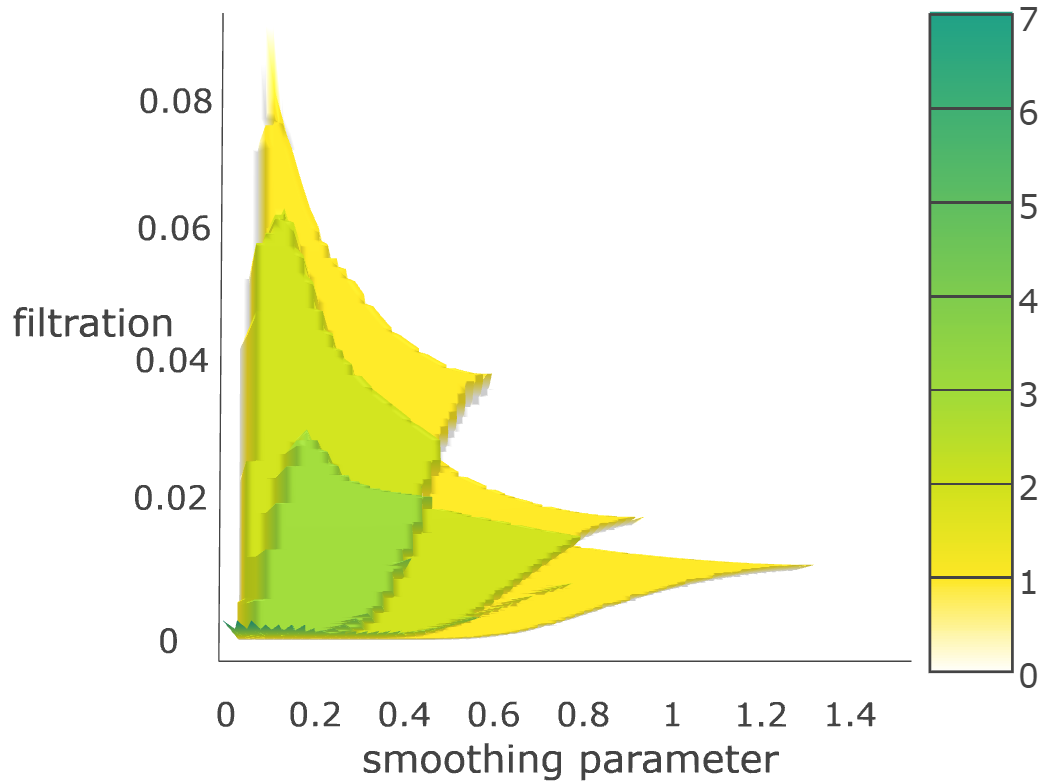}
		\caption{75 smoothing parameters}\label{fig:pt75}
	\end{subfigure}
	\begin{subfigure}[b]{0.49\textwidth}
		\centering
		\includegraphics[width=1\textwidth]{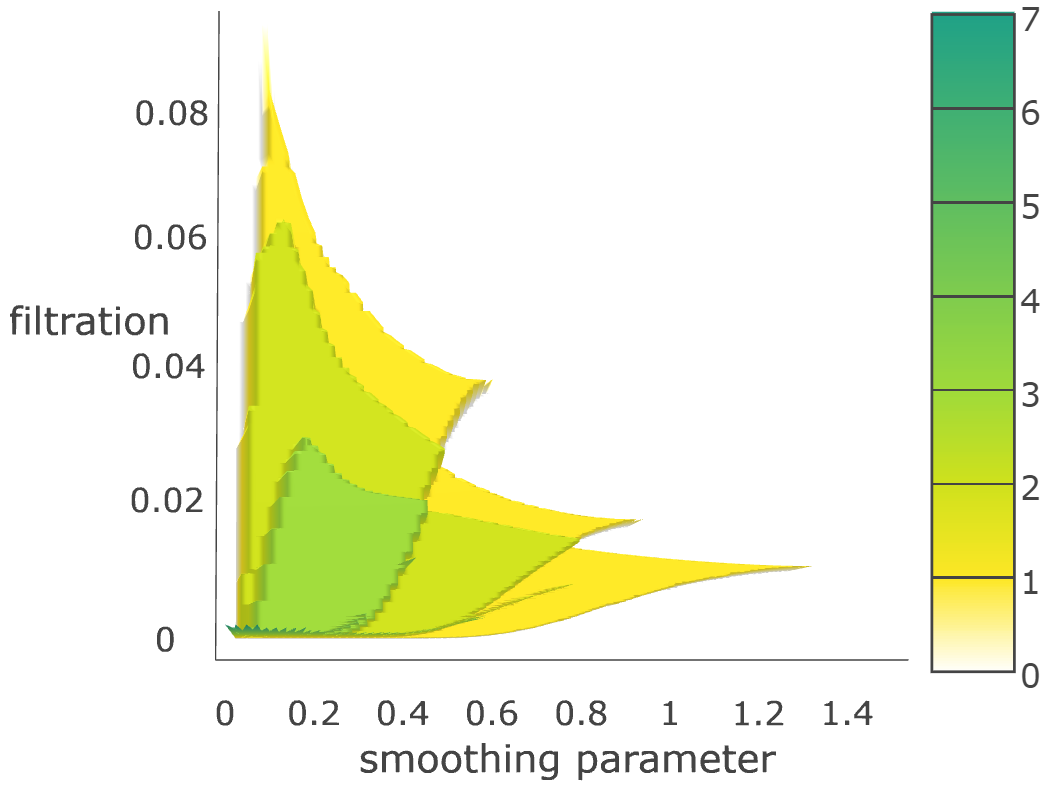}
		\caption{100 smoothing parameters}\label{fig:pt100}
	\end{subfigure}
	\caption{Resolution of the persistence terrace according to the number of smoothing parameters.  }
	\label{fig:resolution}
\end{figure}
With an increase in the number of smoothing parameters, the resolution increases, although the general picture stays the same.